\documentclass[useAMS,usenatbib]{mnras}

\pdfoutput=1

\usepackage[pdftex]{graphicx,color}
\usepackage[T1]{fontenc}

\definecolor{urlblue}{rgb}{0,0,0.9}
\definecolor{linkblue}{rgb}{0,0,.8}
\definecolor{linkgreen}{rgb}{0,0.45,0}
\definecolor{linkpurple}{rgb}{0.7,0.0,0.4}
\definecolor{linkorange}{rgb}{0.7,0.1,0.0}

\usepackage{amsmath, amssymb}
\usepackage{cuted}
\usepackage[utf8]{inputenc}
\usepackage{lmodern}
\usepackage[english]{babel}

\AtBeginDocument{\hypersetup{
linkcolor=linkblue,
citecolor=linkorange,
urlcolor=urlblue}}

\providecommand{\eprint}[1]{\href{http://arxiv.org/abs/#1}{#1}}
\providecommand{\adsurl}[1]{\href{#1}{ADS}}

\usepackage{ifthen}\def\eprinttmp@#1arXiv:#2 [#3]#4@{\ifthenelse{\equal{#3}{x}}{\href{http://arxiv.org/abs/#1}{#1}}{\href{http://arxiv.org/abs/#2}{arXiv:#2} [#3]}}
\renewcommand{\eprint}[1]{\eprinttmp@#1arXiv: [x]@}

\newcommand{\sint}{\sigma_{\rm int}}

\newcommand{\dd}{\textrm{d}}
\def\rd{{\rm d}}

\newcommand{\tgl}{\texttt{turboGL}\ }

\newcommand{\threeint}{\mu_{3,\rm int}}
\newcommand{\fourint}{\mu_{4,\rm int}}

\newcommand{\jpas}{\mbox{J-PAS}\ }

\usepackage{etoolbox}
\makeatletter
\newcount\c@additionalboxlevel
\newcount\c@maxboxlevel
\patchcmd\@combinedblfloats{\box\@outputbox}{%
  \stepcounter{additionalboxlevel}%
  \box\@outputbox
}{}{\errmessage{\noexpand\@combinedblfloats could not be patched}}

\AtBeginShipout{%
  \ifnum\value{additionalboxlevel}>\value{maxboxlevel}%
    \typeout{Warning: maxboxlevel might be too small, increase to %
      \the\value{additionalboxlevel}%
    }%
  \fi
  \@whilenum\value{additionalboxlevel}<\value{maxboxlevel}\do{%
    \typeout{* Additional boxing of page `\thepage'}%
    \setbox\AtBeginShipoutBox=\hbox{\copy\AtBeginShipoutBox}%
    \stepcounter{additionalboxlevel}%
  }%
  \setcounter{additionalboxlevel}{0}%
}
\makeatother

\title[Constraining the HMF with observations]{Constraining the halo mass function with observations}

\author[Castro, Marra and Quartin]{
Tiago Castro,$^{1,2}$
Valerio Marra$^{3}$
and Miguel Quartin$^{1}$
\\
$^{1}$Instituto de Física, Universidade Federal do Rio de Janeiro, 21941-972, Rio de Janeiro, RJ, Brazil\\
$^{2}$Dipartimento di Fisica, Sezione di Astronomia, Università di Trieste, Via Tiepolo 11, I-34143 Trieste, Italy\\
$^{3}$Departamento de Física, Universidade Federal do Esp\'{\i}rito Santo, 29075-910, Vit\'oria, ES, Brazil
}

\date{Accepted 2016 August 16. Received 2016 July 29; in original form 2016 May 30}

\pubyear{2016}

\begin{document}
\label{firstpage}
\pagerange{\pageref{firstpage}--\pageref{lastpage}}

\maketitle

\begin{abstract}
The abundances of dark matter halos in the universe are described by the halo mass function (HMF). It enters most cosmological analyses and parametrizes how the linear growth of primordial perturbations is connected to these abundances. Interestingly, this connection can be made approximately cosmology independent. This made it possible to map in detail its near-universal behavior through large-scale simulations. However, such simulations may suffer from systematic effects, especially if baryonic physics is included. In this paper we ask how well observations can constrain directly the HMF. The observables we consider are galaxy cluster number counts, galaxy cluster power spectrum and lensing of type Ia supernovae. Our results show that DES is capable of putting the first meaningful constraints on the HMF, while both Euclid and \jpas can give stronger constraints, comparable to the ones from state-of-the-art simulations. We also find that an independent measurement of cluster masses is even more important for measuring the HMF than for constraining the cosmological parameters, and can vastly improve the determination of the halo mass function. Measuring the HMF could thus be used to cross-check simulations and their implementation of baryon physics. It could even, if deviations cannot be accounted for, hint at new physics.
\end{abstract}


\section{Introduction}\label{intro}

Dark matter halos play an important role in cosmology as they model galaxies and galaxy clusters which are essential in the study of the large-scale structure of the universe.
Therefore, the halo mass function, which describes how many halos of a given mass exist at a given redshift, is central in cosmology.
The halo mass function allows us to understand the statistics of primordial matter inhomogeneities and is also necessary to compute the effect of nonlinear structures on observations through, for instance, the Sunyaev-Zeldovich effect and lensing.
A very interesting feature of the halo mass function is that -- at least within the standard model of cosmology  -- it acquires an approximate universality when expressed with respect to the variance of the mass fluctuations.

\citet{Press:1973iz} provided the first quantitative model for this universal function, later followed by a wealth of theoretical studies guided by ever better simulations of structure formation~\citep{Sheth:1999mn,Jenkins:2000bv,White:2002at,Springel:2005nw,Warren:2005ey}.
Analytical fits to the simulated mass functions have been obtained in an ever wider mass range~\citep{Kim:2011ab,Angulo:2012ep,Fosalba:2013wxa}, and the possible impact of the evolution of the dark energy has been studied~\citep{Courtin:2010gx,DeBoni:2010nz,Baldi:2011qi}, also in connection with a possible violation of its universality~\citep{Tinker:2008ff}. In recent years, numerical codes that include the effect of baryons have been developed~\citep[][]{Teyssier:2001cp,Wadsley:2003vm,Springel:2009aa,2013MNRAS.431.1366S,Hopkins:2014qka,Bocquet:2015pva} and  corrections to pure-dark matter mass functions have been proposed~\citep{Cui:2014aga,Velliscig:2014bza,Vogelsberger:2014dza}.
All the above work plus the studies of the effect of primordial non-Gaussianities on halo abundances~\citep{Grossi:2009an,Giannantonio:2009ak,Wagner:2010me,LoVerde:2011iz} show how great an effort has been put in the study of the halo mass function.

Cosmological data analysis heavily relies on the latter results in order to infer physical quantities relevant for structure formation such as the matter density parameter, the power spectrum normalization or the growth rate index. Or to compute, for instance, the expected small-scale nonlinear power spectrum~\citep[see][for a revised halo model]{Mead:2015yca},%
\footnote{The halo model has also been used to compute the nonlinear bispectrum~\citep{2000ApJ...543..503M}.}
necessary to compute for example lensing corrections to observables such as the Cosmic Microwave Background.

However, while individual state-of-the-art simulations may have a very high precision, that is, small statistical errors on the parameters that enter the mass function, their accuracy may not necessarily be so high~\citep{Knebe:2013xwz,Murray:2013sna,Casarini:2014rua}.
Indeed, many are the possible causes of systematic differences between simulations which could affect cosmological forecasts~\citep{Cunha:2009rx,Paranjape:2014lga,Schneider:2015yka,Bocquet:2015pva}. Some properties such as resolution, box size, halo finder and implementation of initial conditions are more straightforward to check. Others, like the implementation of baryon physics at many different levels, including cooling mechanisms, both star and galaxy formation, supernova and AGN feedback, and the overall hydrodynamic approximation used, are decided at a much more \emph{holistic} level -- see for instance~\citet{Crain:2015poa}.
In other words, a large uncertainty both in the different individual baryonic effects and on their interplay remains. Simulation parameters (including non-simulated sub-grid physical effects) are thus adjusted in order to have an overall good fit to astronomical observations.
For example, different baryonic implementations were thoroughly tested in \citet{Sembolini:2015mla,Sembolini:2015dhr,2016MNRAS.458.1096E, 2016MNRAS.458.4052C} where different $N$-body codes have been compared, showing that stronger inconsistencies appear for halos defined according to a higher mean density.
Therefore, it is interesting to see if the halo mass function can be constrained/reconstructed directly from observations: this can be used as a further cross-check on the numerical simulations.
Furthermore, there may be physics beyond the standard model which plays an important role in structure formation. Violation of statistical isotropy, clustering dark energy, modified gravity, nonstandard primordial fluctuations could, for instance, significantly alter halo abundances.

In this paper we explore how observations can constrain the mass function. We focus on three observables. The first is galaxy cluster counts which is the most obvious (and probably the most powerful) observable to consider as it directly constrains halo abundances. We consider the forecasted cluster catalogs from the Euclid Mission~\citep{Laureijs:2011gra}, from the Javalambre-Physics of the Accelerated Universe Astrophysical Survey~\citep[\jpas --][]{Benitez:2014ibt} and  from the Dark Energy Survey~\citep[DES --][]{Abbott:2005bi}. The second observable is the galaxy cluster power spectrum (usually considered in \emph{halo self-calibration} techniques~\citep{Majumdar:2002hd,Majumdar:2003mw,Lima:2004wn,Lima:2005tt}), for which one can use the same data we use for the galaxy cluster counts. The third observable is supernova Ia lensing, which has been recently adopted in~\citet{Quartin:2013moa,Castro:2014oja,Amendola:2014yca,Castro:2015rrx}. Other observables such as  lensing of galaxies (see~\citealt{Troxel:2014dba} for a review) should also be able to constrain the halo mass function. However, as we show below, the three probes here considered are almost orthogonal in the mass function parameter space and thus their combination alone is already able to give constraints on the mass function parameters that are comparable to the statistical errors from state-of-the-art simulations. In particular, future Euclid and \jpas data could, at the same time, confirm results from simulations or hint for physics beyond the standard model, if deviations will be detected.

This paper is organized as follows. In Section~\ref{sec:hmf} we introduce halo mass function and bias and how we parametrize them.
In Section~\ref{sec:sys} we show the systematic differences in mass functions from different simulations.
In Section~\ref{meda} we build the likelihood functions for the cluster number counts, cluster power spectrum and supernova lensing observables, while in Section~\ref{results} we show how these observables can constrain the mass function with forecasted catalogs.
We draw our conclusions in Section~\ref{conclusions}. Finally, in Appendix~\ref{app:pfits} fitting functions for the lensing moments as a function of redshift and the halo mass function parameters are given.

\section{The halo mass function and bias} \label{sec:hmf}

The halo mass function $f(M,z)$ gives the fraction of the total mass in halos of mass $M$ at redshift $z$. It is related to the (comoving) number density $n(M,z)$ by:
\begin{equation} \label{hmf}
    \rd n(M,z) \,\equiv\, n(M,z) \rd M \,=\, {\rho_{mc} \over M} \, f(M,z) \rd M \,,
\end{equation}
where $\rho_{mc}$ is the constant matter density in a co-moving volume, and we defined $\rd n$ as the number density of halos in the mass range $\rd M$. The halo function is by definition normalized to unity
\begin{equation}
    \int f(M,z) {\rm d}M \,=\, 1 \,.
\end{equation}

This function acquires an approximate universality when expressed with respect to the variance of the mass fluctuations on a comoving scale $r$ at a given redshift $z$, $\Delta(r,z)$.
Relating the comoving scale $r$ to the mass scale by $M = (4 \pi / 3) \, r^{3}  \, \rho_{mc}$, we can define the variance in a given mass scale by
\begin{equation}
    \Delta^{2}(M,z) \,\equiv\, \Delta^{2}(r(M),z) \,.
\end{equation}
The variance $\Delta^2(r,z)$ can be computed from the power spectrum:
\begin{equation}
    \Delta^{2}(r,z) \,=\,\int_{0}^{\infty} {dk \over k} \tilde{\Delta}^{2}(k,z) W^{2}(k  r) \, ,
\end{equation}
where $W(k  r)$ is the Fourier transform of the top-hat window function and $\tilde{\Delta}^{2}(k,z)$ is the dimensionless power spectrum extrapolated using linear theory to the redshift $z$:
\begin{equation} \label{linearpk}
    \tilde{\Delta}^{2}(k,z) \equiv {k^{3} \over 2 \pi^{2}} P(k, z) =
    \delta_{H0}^{2} \left( {c k \over H_{0}} \right )^{3+n_{s}} T^{2}(k)\, G^{2}(z) \, .
\end{equation}
In the previous equation $P(k, z)$ is the power spectrum, $n_{s}$ is the spectral index, $\delta_{H0}$ is the amplitude of perturbations on the horizon scale today, $G(z)$ is the linear growth function~\citep[see, e.g.,][]{Percival:2005vm}, and $T(k)$ is the transfer function~\cite[see][equations (28)--(31)]{Eisenstein:1997ik}.

With $\Delta\equiv\Delta(M,z)$ given, we can now define our halo mass function, which we model according to the functional form proposed by~\citet{Sheth:1999mn}:
\begin{equation} \label{st}
    f_{\rm ST}(\Delta) \,=\, A\sqrt{{2a\over\pi}} \left[1+\left({\Delta^2\over a\delta_c^2}\right)^p\right] {\delta_c\over\Delta}\exp\left[-{a\delta_c^2\over2\Delta^2}\right],
\end{equation}
where $\,a>0\,$ and $\,\delta_c\,$ is the linear-theory critical threshold for collapse at $z=0$, which we model according to \citet{Weinberg:2002rd}.
Including the cosmology-dependent $\,\delta_c\,$ in the analysis is one of the important advantages~\citep{Courtin:2010gx} of the Sheth-Tormen template with respect to templates based solely on $\Delta^2$~\citep[see e.g.][]{Tinker:2008ff}.
Note that, because of the change of variable, $f_{\rm ST}$ is related to our original  definition of $f$ by:
\begin{equation}
    f(M,z) \,=\, f_{\rm ST}(M,z) {\dd\ln \Delta(M,z)^{-1} \over \dd M} \, .
\end{equation}

In equation \eqref{st} $\,a\,$ and $\,p\,$ are free parameters, and $\,A\,$ is fixed by the normalization $\,\int f \dd M=1$. The integral can be performed analytically, yielding the following useful relation:
\begin{equation} \label{Aofp}
    A\,=\,A(p)\,=\,\left[ 1+ \frac{2^{-p}}{\sqrt{\pi}} \Gamma(1/2-p) \right]^{-1} \,,
\end{equation}
where $p<1/2$. The fiducial values we will adopt are the original ones of~\citet{Sheth:1999mn}:
\begin{equation} \label{stfid}
    \{a_{\rm fid}, p_{\rm fid} \} \,=\,\{ 0.707, 0.3 \} \,.
\end{equation}
The mass function we consider
has recently been outperformed in precision by more recent templates with more parameters, especially as far as the high-mass tail of the mass function is concerned \citep{Tinker:2008ff,Bocquet:2015pva,Despali:2015yla}. For instance, \citet{Despali:2015yla} left the normalization $\,A\,$ free in order to account for our poor knowledge of $f(M,z)$ outside the mass range constrained by the simulations, while \citet{Tinker:2008ff,Bocquet:2015pva} used redshift dependent parameters in order to ensure the universality of their fitting functions. In this first analysis  we nevertheless focus on the simpler template~\eqref{st} as the two parameters $\,a\,$ and $\,p\,$ already capture most of the information available in the mass function and the additional parameters that refine more recent templates are more difficult to constrain.

Halos are biased tracers of the underlying dark matter field. Using the peak-background split it is possible to obtain the halo bias $b(M)$ directly from the halo mass function~\citep{Sheth:1999mn}:
\begin{equation} \label{bias}
b(M) = 1 + \frac{{a\delta_c^2}/{\Delta^2}-1}{\delta_c}
+ \frac{2p}{\delta_c [1+({a\delta_c^2}/{\Delta^2})^p]} \,.
\end{equation}
As we will see below, the knowledge of the halo bias allows one to calculate the large-scale clustering of halos.

The definition of a mass function is always accompanied by a prescription on how to define the halos and their masses. This prescription affects the values of $\,a\,$ and $\,p\,$. Here, we model the halos as spherical structures. The physical radius $R_{p}$ of a halo of mass $M$ is defined so that the halo average density is $\Delta_{\rm SO}$ times the background matter density $\rho_{m}(z)$ or the critical density $\rho_{\rm crit}(z)$:%
\footnote{Both definitions are common in the literature although both share the undesired property of being cosmology dependent through a fixed overdensity with respect to the mean or critical density. See also \citet{Despali:2015yla} where the use of the virial overdensity is shown to improve the universality of the mass function for the case of the standard $\Lambda$CDM model.}
\begin{equation}  \label{halorad}
    M \,=\, {4 \pi \over 3}  R_{p}^{3}   \, \rho_{m/{\rm crit}}(z) \, \Delta_{\rm SO} \, .
\end{equation}
In other words, halos are defined according to a spherical-overdensity (SO) halo finder as it is now common in $N$-body simulations \citep[see e.g.][]{Tinker:2008ff}. In the following halo masses are defined according to $\Delta_{\rm SO}=200$ with respect to $\rho_{\rm crit}(z)$.

Summarizing, $a$ and $p$ parametrize how the linear growth of perturbations $G(z)$ is connected to the halo abundances $f(M,z)$ and the halo bias $b(M)$.
This connection should be approximately cosmology independent, a property inherited by the mass function.
Furthermore, it should be approximately independent of the nonlinear (possibly baryonic) physical processes leading to the halo density profiles, as long as the halo masses as defined above remain unchanged~\citep[see, however,][]{Cui:2014aga,Velliscig:2014bza,Vogelsberger:2014dza}.

\section{Systematic differences in halo mass function parameters} \label{sec:sys}

In this Section we aim at substantiating the claim made in the Introduction, to wit that halo mass functions from simulations may suffer from systematic differences which are larger than their statistical uncertainties.

As said earlier, we consider halo masses defined according to $\Delta_{\rm SO}=200$ with respect to~$\rho_{\rm crit}(z)$.
\citet{Despali:2015yla} is the only recent paper that uses the Sheth-Tormen mass function of equation~\eqref{st} with SO masses. Therefore, in order to carry out this investigation, we fitted the ST template to three recent $N$-body simulations:
the hydrodynamical \citep{Bocquet:2015pva} and dark-matter (Castro et al., in prep.)%
\footnote{In this paper we have used the dark matter counterpart of the Box0 simulation presented in \citet{Bocquet:2015pva}. More details of this simulation will be presented in a future paper about halo bias.} Magneticum simulations and the Millennium-XXL simulation of \citet{Angulo:2012ep}.%
\footnote{We used the publicly available halo catalog which contains halos more massive than $10^{13} h^{-1} M_{\odot}$.}
In these simulations halos were found using the \texttt{SUBFIND} \citep{Springel:2000qu,Dolag:2008ar} algorithm which detects gravitationally bound structures in parent groups that were previously identified by a Friends-of-Friends (FoF) algorithm. In the Millennium simulation a FoF linking length $b=0.2$ was adopted, while in the Magneticum simulation a linking length $b=0.16$ was used.

\begin{figure}
\begin{centering}
\vspace{-0.5cm}
\includegraphics[width=.9\columnwidth]{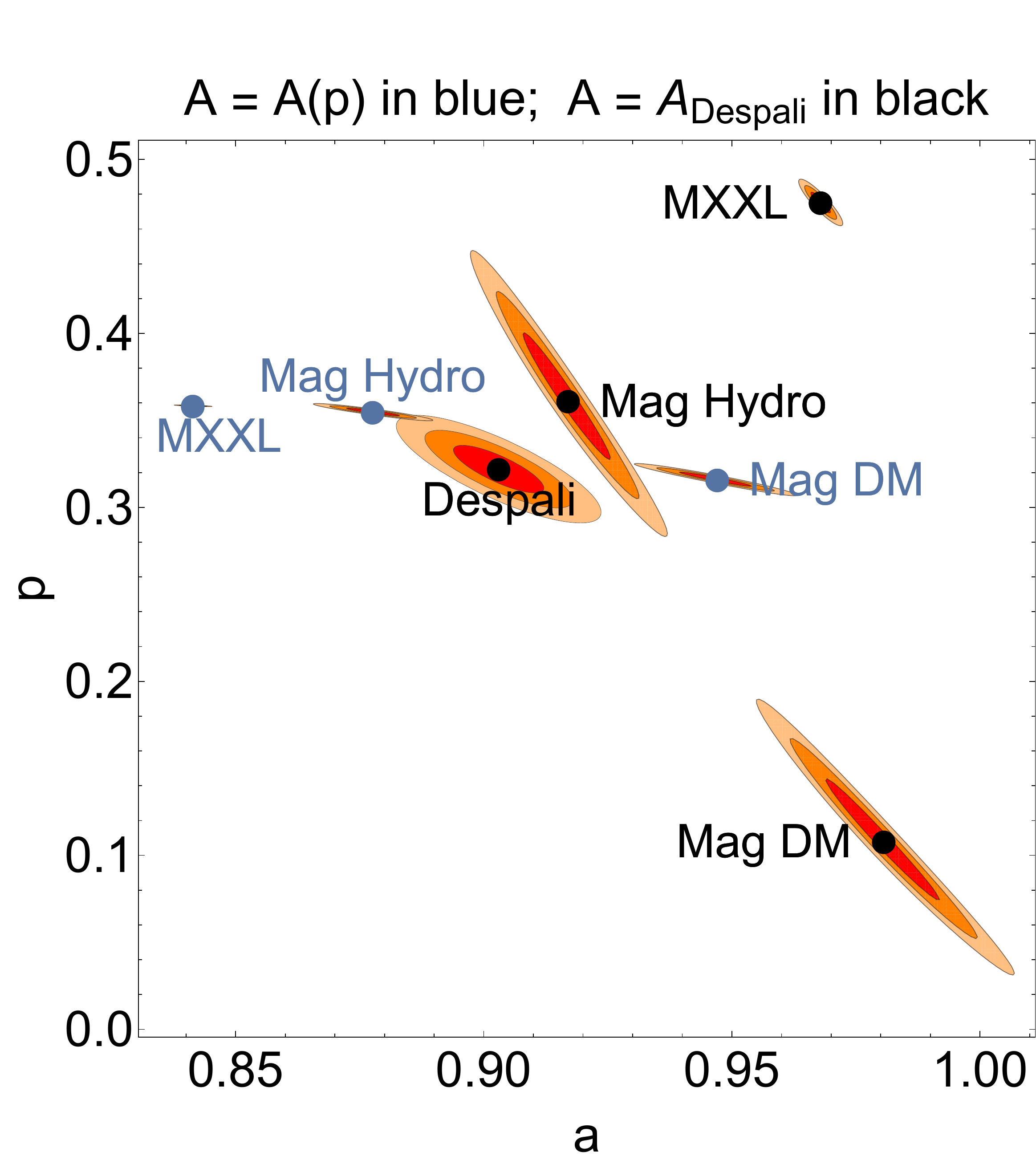}
\caption{
Parameters $a$ and $p$ of the Sheth-Tormen mass function in~\eqref{st} obtained from different simulations. \emph{Blue:} the normalization parameter $A$ is fixed by demanding overall normalization $A(p)$ of eq.~\eqref{Aofp}. \emph{Black:} $A=A_{\rm Despali}$. `Mag DM' (`Mag Hydro') refers to the pure dark-matter (fully hydrodynamical) simulations in \citet{Bocquet:2015pva};
`Despali', to DM simulations in~\citet{Despali:2015yla}; `MXXL',  to DM simulations in~\citet{Angulo:2012ep}.
Note that the statistical (Poissonian) errors -- depicted with 1-, 2- and 3$\sigma$ contours -- are  always smaller than systematic differences in the various simulations. Halo masses are defined according to $\Delta_{\rm SO}=200$ with respect to~$\rho_{\rm crit}(z)$. See Section~\ref{sec:sys} for more details.
}
\label{stplot}
\end{centering}
\end{figure}

In the ST template, the normalization parameter $A$ can be fixed (as it is in the following sections) by demanding overall normalization with \eqref{Aofp} (blue circles in Figure~\ref{stplot}) or can be left free in order to improve the performance of the Sheth-Tormen mass function on the relevant mass range.
\citet{Despali:2015yla} considers only this second choice. Therefore, in order to compare with their results we also consider the case of $A=A_{\rm Despali}$ (black circles in Figure~\ref{stplot}).

Figure~\ref{stplot} shows the different best-fit parameters for the various cases. The statistical (Poissonian) errors -- depicted with 1-, 2- and 3$\sigma$ contours -- are clearly always smaller than systematic differences between the parameters inferred from the various simulations. Differences between pure-DM simulations such as `Despali',%
\footnote{For our choice ($\Delta_{\rm SO}=200$ with respect to critical density) the authors quote the errors but not the correlation between $a$ and $p$. We have thus assumed the same correlation shown in their Fig.~9 (to wit, $-0.8$), which uses the virial overdensity and the redshift range $0<z<1.25$.}
`MXXL' and `Mag DM' are expected to come from different resolutions, box sizes, halo finders and implementations of initial conditions. The difference between `Mag DM' and `Mag Hydro' is instead entirely due the particular implementation of baryon physics.

It is important to stress that the results shown in Figure~\ref{stplot} may depend on the chosen mass function template as its number of free parameters may affect the statistical uncertainties and thus the significance of the systematic differences.
Also, the best-fits shown in Figure~\ref{stplot} are, to some extent, affected by the not optimal performance of the Sheth-Tormen mass function for high halo masses. This is clearly shown by the contours relative to the MXXL simulation, where the very large number of smaller halos reduces the poissonian variance to such a level that the best-fit values of $a$ and $p$ move by many sigmas once the parameter $A$ is changed to the value used in \citet{Despali:2015yla}.
Nevertheless, it is important to note that such a difference results in a HMF that differs by less than 10\% in the mass range of calibration.

\section{Method and data}\label{meda}

\subsection{Galaxy cluster number counts}
\label{clusters1}

Galaxy cluster number counts is the most obvious observable to consider as it directly constrains the halo abundances $\rd n(M,z)$ and so the halo mass function of equation~\eqref{hmf}.
However, rather than constraining the mass function itself, number counts have been used so far to constrain the properties of the linear density field $\Delta (M,z)$, in particular the matter density parameter $\Omega_{m0}$ and the power spectrum normalization $\sigma_8$~\citep[see, e.g.,][and references therein]{Ade:2015fva}. The reason behind this choice is that it is believed that the halo mass function can be determined with high enough precision from sufficiently sophisticated large-scale simulations. In other words, the systematic error in the mass function is believed to be subdominant compared to the other sources of error that are involved in the determination of, say, $\Omega_{m0}$ and $\sigma_8$.

Here, as argued in the Introduction, we explore the inverse approach. We fix the cosmological parameters to the best-fit values from Planck~\citep[][table 4, last column]{Ade:2015xua} and use forecasted number counts in order to constrain the Sheth-Tormen mass function parameters $a$ and $p$ of equation~\eqref{st}. Clearly, an ideal data analysis should take into account simultaneously \emph{both} the uncertainties in the cosmology and in the mass function. However as mentioned above, fixing the cosmology can be justified because the mass function -- and so its parameters $\,a\,$ and $\,p\,$ -- should be approximately cosmology independent. Moreover, it makes our present analysis much simpler.

The number of clusters expected in a survey with sky coverage $\Omega_{\rm sky}$ within the $i$-th redshift bin $\Delta z_i$ centered around $z_i$ and the $j$-th mass bin $\Delta M_j=M_{j+1}-M_j$ is \citep[see, e.g.,][]{Sartoris:2015aga}:
\begin{align} \label{Nij}
    N_{ij} \,=\;\, &\frac{\Omega_{\rm sky}}{8 \pi} \int_{\Delta z_i} \rd z \frac{\rd V}{\rd z} \nonumber \\
    &\int_0^{\infty} \rd M \, n(M, z) \, \left( \textrm{erfc} \, x_j - \textrm{erfc} \, x_{j+1}  \right) \,,
\end{align}
where the lowest mass bin corresponds to $M_{{\rm thr},i}$ which is the limiting cluster mass at redshift $z_i$ that the survey can detect at a given signal-to-noise ratio (see Figure~\ref{mthr}). The quantity $\rd V/\rd z$ is the cosmology-dependent comoving volume element per unit redshift interval which is given by:
\begin{equation} \label{volume}
    \frac{\rd V}{\rd z} \,=\, 4 \pi  (1+z)^2 \, \frac{d_A^2(z)}{c^{-1} H(z)} \,,
\end{equation}
where $d_A$ is the angular diameter distance and $H(z)$ is the Hubble rate at redshift $z$.

In equation~\eqref{Nij} $\textrm{erfc}(x)$ is the complementary error function and $x$ is:
\begin{equation}
    x(M_{\rm ob})\,=\, \frac{\ln M_{\rm ob}-\ln M_{\rm bias}-\ln M}{\sqrt{2} \sigma_{\ln M}} \,,
\end{equation}
where $M_{\rm bias}$ models a possible bias in the mass estimation and $\sigma_{\ln M}$ is the intrinsic scatter in the relation between true and observed mass.
We model the latter two quantities as in~\citet{Sartoris:2015aga}:
\begin{align}
\ln M_{\rm bias} &\,=\, B_{M0} + \alpha \ln (1+z) \,, \label{Mbias} \\
\sigma_{\text{ln}M}^2 &\,=\, \sigma_{\text{ln}M0}^2  + (1+z)^{2 \beta} -1\,. \label{sigln}
\end{align}
It is important to point out that while the parameters $a$ and $p$ are approximately cosmological independent, the nuisance parameters may not be. Here, we neglect this possible dependence or, equivalently, we assume that cosmology has been sufficiently well constrained by other data.

We then assume Poisson errors for the cluster counts so that we can use the Cash $C$ statistics \citep{Cash:1979vz,Holder:2001db}:
\begin{equation} \label{Lcc}
C \,=\, - 2 \ln L_{\rm cc} \,=\, 2 \sum_{ij} \left (N_{ij} - N_{ij}^{\rm obs} \ln N_{ij} \right) ,
\end{equation}
where $L_{\rm cc}$ is the cluster count likelihood, and $N_{ij}$ and $N_{ij}^{\rm obs}$ are expected and observed counts, respectively.
Equation~\eqref{Lcc} is valid if the bins are uncorrelated; in other words, correlations due to large-scale clustering are neglected. This should be a good approximation as the cluster catalogs we consider contain very massive systems for which the impact of correlations is negligible~\citep[see][]{Hu:2002we}.%
\footnote{%
As a side note, we would like to point out that we obtain basically unchanged results if instead of $C$ we adopt its Gaussian approximation $C^{\rm gauss}= \sum_{ij} \left ( N_{ij} - N_{ij}^{\rm obs} \right )^2/N_{ij}$.
}
Summarizing, $L_{\rm cc}$ depends on the six parameters~$\{a, \, p,\,B_{M0}, \,\alpha,\,\sigma_{\text{ln}M0}, \,\beta  \}$.

It is important to note that the analysis of real cluster data is more complicated than the straightforward approach of this section. This is due to the fact that in any cluster catalog one needs to put a large effort in understanding both its completeness and purity~\citep[see e.g.][]{Bleem:2014iim,Ade:2015gva}. The impurity of a catalog, if not well understood, can lead to biases on the results, and most catalogs used for cosmology rely on high-purity quality cuts. The incompleteness of a catalog, on the other hand, is usually corrected for by estimating for each observed cluster the corresponding effective completeness. This is nevertheless not a straightforward task, as the completeness depends on the observed position, mass, size and signal-to-noise (S/N) of the given cluster, as well as on the assumed cosmology, as discussed in~\citet{Ade:2015fva}. In fact, in that paper the Planck collaboration does not even analyze their data separately in different mass bins. Here we avoid these important complications by dealing only with forecasted data, which we discuss next.

\subsubsection{Cluster counts data}
\label{clusters2}

\begin{figure}
\begin{centering}
\includegraphics[width=.99\columnwidth]{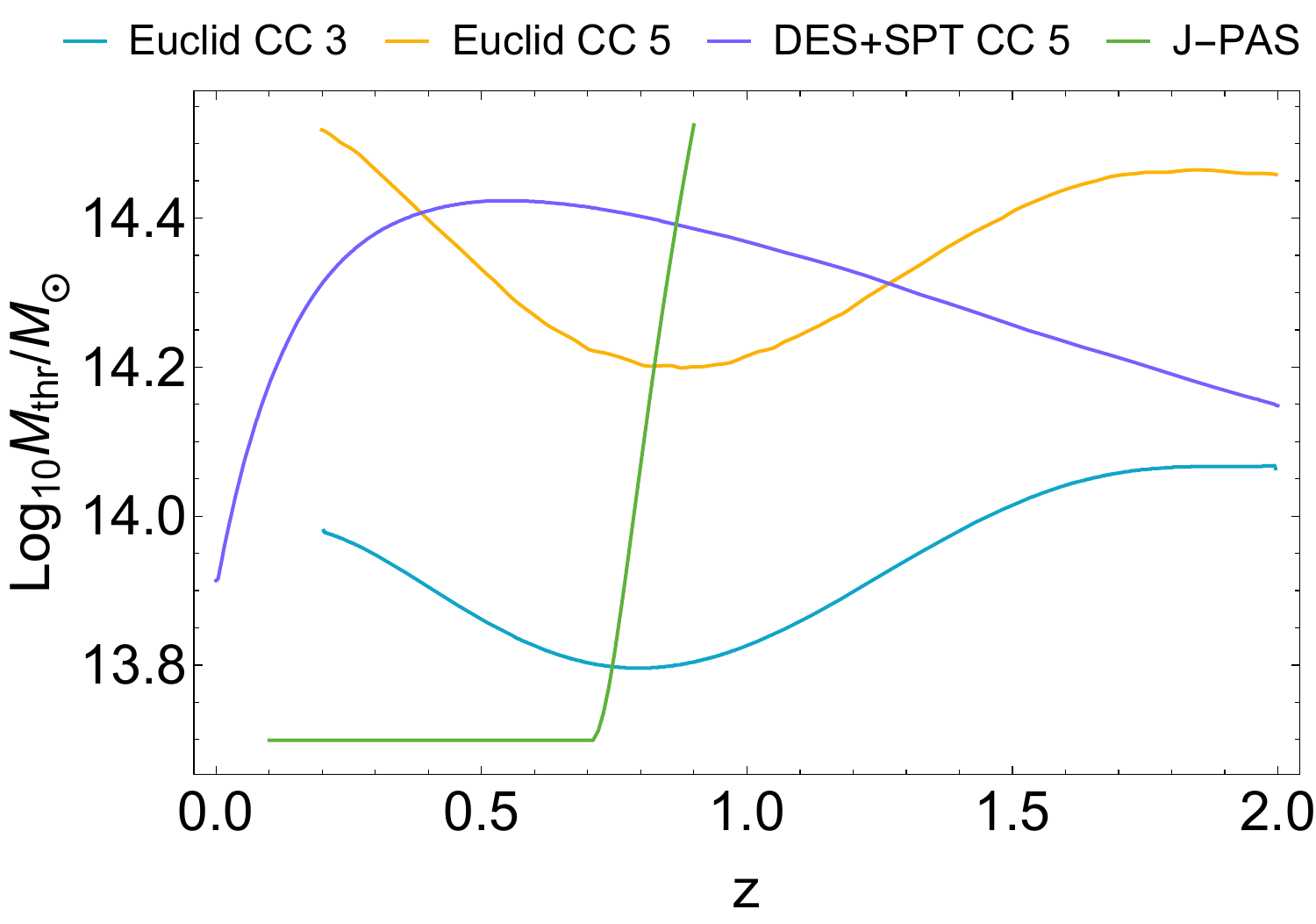}
\caption{Mass-threshold value as a function of redshift of the observed cluster mass for Euclid for a detection threshold of 3 (Euclid CC 3, in the plot) and 5 (Euclid CC 5)~\citep[see][Fig.~2]{Sartoris:2015aga}, for the Javalambre-Physics of the Accelerated Universe Astrophysical Survey (J-PAS)~\citep[see][Fig.~12]{Ascaso:2016ddl},
and for the Dark Energy Survey for a detection threshold of 5 (DES+SPT CC 5)~\citep[see][Fig.~2.1]{Abbott:2005bi}.
See Section~\ref{clusters2} for more details.}
\label{mthr}
\end{centering}
\end{figure}

As mentioned in the Introduction, we consider number counts forecasts from Euclid, \jpas and DES. All number count forecasts need to assume some quality cuts criteria or S/N thresholds in order to ensure a sufficient complete and pure catalog. Both Euclid and DES parametrize this choice in a S/N threshold defined as the ratio between the number of galaxies associated to a cluster and the root mean square of field counts.

The Euclid spacecraft is currently under construction and scheduled for launch in 2020. During its mission, which will last at least 6 years, Euclid will observe approximately $\Omega_{\rm sky}=15000 \deg^2$ of the extragalactic sky, which is about half of the total sky facing away from the Milky Way. Following~\citet{Sartoris:2015aga}, its limiting cluster mass is shown in Figure~\ref{mthr} for a detection S/N threshold of 3 and 5. A detection threshold of 3 roughly corresponds to 80\% completeness. The shape of the selection functions is higher at $z\sim0.2$ than at $z\sim0.7$, while one would expect the opposite (as it is the case for the DES selection function discussed below). \citet{Sartoris:2015aga} explains this counter-intuitive behavior as due to the relative importance of cosmic variance and Poisson noise in the contaminating field counts.
As in \citet{Sartoris:2015aga}, the fiducial values for the four nuisance parameters of equations \eqref{Mbias}--\eqref{sigln} are:
\begin{equation}
\begin{aligned}
\{B_{M0,\rm{fid}}, \;\alpha_{\rm{fid}} \}&\;=\;\{ 0, \; 0 \} \,,  \\
\{ \sigma_{\text{ln}M0,\rm{fid}}, \;\beta_{\rm{fid}}\}&\;=\;\{ 0.2, \, 0.125 \} \,.
\label{nuilab}
\end{aligned}
\end{equation}
In \citet{Sartoris:2015aga} the halo masses have been defined according to $\Delta_{\rm SO}=200$ with respect to $\rho_{\rm crit}(z)$ --~see equation \eqref{halorad}; consequently, the constraints on the mass function parameters we obtain in Section~\ref{results} are relative to this definition.
Figure~\ref{mthr} also shows the redshift range covered by the Euclid mission.

\jpas is a ground-based survey that is expected to begin scientific observations by the end of 2016. It will observe approximately $8500 \deg^2$ of the northern sky with 54 narrow-band filters plus two medium-band and three broad-band filters in the whole optical range. Thanks to its quasi-spectroscopic photometric redshift \jpas provides near optimal efficiency for separating cluster members from foreground and background galaxies.
Indeed, the accuracy of the photometric redshift matches the typical velocity dispersion of massive clusters, therefore allowing to detect clusters above the noise to much lower masses and higher redshifts than wide-field surveys using conventional filters.
Its low limiting cluster mass is shown in Figure~\ref{mthr} and corresponds to $>80\%$ of completeness and purity~\citep{Ascaso:2016ddl}.
In this case the selection function has been approximated as constant for $z \le 0.7$.
The fiducial values of the J-PAS nuisance parameters are:
\begin{align}
\{B_{M0,\rm{fid}}, \;\alpha_{\rm{fid}} \}&\;=\;\{ 0, \; 0 \} \,, \nonumber \\
\{ \sigma_{\text{ln}M0,\rm{fid}}, \;\beta_{\rm{fid}}\}&\;=\;\{ 0.142, \, 0 \} \,.
\label{nuijpas}
\end{align}

DES is an ongoing survey which started operations in 2013. Regarding the cluster catalog, DES will work in synergy with the South Pole Telescope~\citep[SPT --][]{Carlstrom:2009um}, which conducted a $2500 \deg^2$
extragalactic Sunyaev-Zel'dovich (SZ) survey~\citep{Bleem:2014iim,deHaan:2016qvy} and detected hundreds of clusters. This whole area will be overlapped by DES, and the SPT SZ map will aid DES in the detection and measurements of its clusters. The limiting cluster mass for a detection S/N threshold of 5 is shown in Figure~\ref{mthr}. The same definition of halo mass has been adopted: $\Delta_{\rm SO}=200 \rho_{\rm crit}$.
Although the fiducial values of eq.~\eqref{nuilab} were obtained by~\citet{Sartoris:2015aga} for the Euclid Mission, they are also compatible with the forecasted $\sigma_{\ln M_{200}}$ for DES, as discussed in~\citet{Rykoff:2011xi}.

The cluster count likelihood is computed in the redshift range of Figure~\ref{mthr} with bins of $\Delta z=0.1$. The observed mass range extends from the lowest mass limit ($M_{\rm thr}$ of Figure~\ref{mthr}) up to $\log M/M_{\odot} \leq16$, with $\Delta \log_{10} M/M_\odot = 0.2$.

\subsection{Cluster power spectrum}
\label{cpk1}

The cluster catalogs discussed in Section~\ref{clusters2} can also be used to calculate the cluster power spectrum~$\Delta^2_{\rm cps}$~\citep{Majumdar:2003mw}.
This is interesting because the theoretical prediction for the cluster power spectrum does not involve the introduction of further parameters or functions.
Indeed, it is sufficient to know the halo bias, which is a direct consequence of the halo function  -- see Eq.~\eqref{bias}. In other words, constraints from the cluster power spectrum allow us to extract more information from cluster data ``for free'' and thus to put tighter constraints on the mass function parameters $a$ and~$p$.

The cluster power spectrum is given by~\citep[see][]{Majumdar:2003mw,Sartoris:2010cr}:
\begin{equation} \label{clpk}
P_{\rm cps} (k, z) = b^2_{\rm eff}(z) \, P(k, z) \,,
\end{equation}
where $P(k, z)$ is the linear power spectrum of Eq.~\eqref{linearpk} and $b_{\rm eff}(z)$ is the linear bias weighted by the mass function:
\begin{equation}
b_{\rm eff}(z) = \frac{1}{\bar n (z)} \int_0^{\infty} \rd M \, n(M, z) \, \textrm{erfc} \{x [M_{\rm thr}(z)]\} \, b(M,z)  \,.
\end{equation}
The normalization factor $\bar n (z)$ is the average number density of objects  included in the survey at the redshift $z$:
\begin{equation}
\bar n (z) \,=\, \int_0^{\infty} \rd M \, n(M, z) \, \textrm{erfc} \{x [M_{\rm thr}(z)]\} \,.
\end{equation}
Note that in equation~\eqref{clpk} redshift space distortions (RSD) have been neglected. RSD are useful as they can constrain the growth factor and so the relevant cosmological parameters. In fact, \citet{Veropalumbo:2015dpi} have shown that clusters can be used effectively to measure these distortions. However, as in the present analysis we are keeping the cosmology fixed, the RSD are not giving extra information and we have neglected their contribution.

The cluster power spectrum of Eq.~\eqref{clpk} is valid for a small redshift interval centered around $z$. Observationally, it is convenient to measure the power spectrum over a wide redshift bin.
The cluster power spectrum averaged over the $i$-th redshift bin $\Delta z_i$ centered around $z_i$ is then~\citep{Majumdar:2003mw}:
\begin{equation}
    \bar P_{\rm cps} (k, z_i) \,=\,
    \frac{\displaystyle\int_{\Delta z_i} \rd z \, \frac{\rd V}{\rd z} \, \bar n^2(z) P_{\rm cps} (k, z)}
    {\displaystyle\int_{\Delta z_i} \rd z \, \frac{\rd V}{\rd z} \, \bar n^2(z) } \,,
\end{equation}
that is, the power spectrum is weighted according to the square of the number density of clusters that are included in the survey at redshift $z$.

As the cluster power spectrum probes linear scales, we can assume uncorrelated Gaussian errors so that we can build the following likelihood:
\begin{equation} \label{likepk}
    - 2 \ln L_{\rm cps} \!=\!
    \sum_{i,j} \!\frac{\big[ \bar P_{\rm cps}(k_j,z_i) -  \hat P_{\rm cps}^{\rm obs}(k_j,z_i)\big]^2}{\sigma^2_P (k_j,z_i)}
    + \ln \sigma^2_P (k_j,z_i)  ,
\end{equation}
where we neglected an inconsequential constant factor and the product runs over the redshift bins $\Delta z_i$ centered around $z_i$ and the wavenumber bins $\Delta k_j$ centered around $k_j$. As in~\citet{Sartoris:2011yr,Sartoris:2015aga} we adopt a constant $\Delta z=0.2$ (see Figure~\ref{mthr} for the redshift range), and $\Delta \log(k \,\text{Mpc}) = 0.1$ with $\{ k_{\rm min}, k_{\rm max} \}=\{10^{-3},0.14 \} \text{Mpc}^{-1}$. The coarser redshift bins should make correlations between adjacent bins negligible and the low value of $k_{\rm max}$ should make nonlinear corrections to the power spectrum negligible. In~\eqref{likepk} the variance is given by~\citep{Scoccimarro:1999kp}:
\begin{equation} \label{sigmaPk}
\frac{\sigma^2_P}{\bar P^2_{\rm cps}} = \frac{(2 \pi)^3}{V_s V_k/2} \Big [ 1 + \frac{1}{\bar n(z) \bar P_{\rm cps}(k,z)} \Big]^2 \,,
\end{equation}
where $V_k$ is the $k$-space volume of the bin, $V_k = 4 \pi k^2 \rd k$, and $V_s$ is the survey volume for the redshift bin $\Delta z_i$, which can be computed using \eqref{volume}, $V_s =\Omega_{\rm sky}(4 \pi)^{-1} \!\int_{\Delta z_i} \rd z \frac{\rd V}{\rd z} $.
As for the forecasts of this paper we are assuming constant (in real space) window function and power spectrum, equation~\eqref{sigmaPk} is in agreement with the optimal weighting scheme of~\citet{Feldman:1993ky}.
Also, equation~\eqref{sigmaPk} neglects any anisotropy in the survey volume.

Clusters sample discretely the underlying matter field and the resulting shot noise has to be accounted for.
In the forecasts of Section~\ref{results} we model the observed power spectrum via $P_{\rm cps}^{\rm obs}=\bar P_{\rm cps}^{\rm fid}+ 1/\bar n^{\rm fid}$. Consequently the estimator for the power spectrum, used in \eqref{likepk}, is:
\begin{equation}
\hat P_{\rm cps}^{\rm obs} = P_{\rm cps}^{\rm obs} - \frac{1}{\bar n }
= \bar P_{\rm cps}^{\rm fid}+ \frac{1}{\bar n^{\rm fid} }- \frac{1}{\bar n } \approx \bar P_{\rm cps}^{\rm fid}   \,,
\end{equation}
where for the last approximation we have used the fact that the mass function parameters will be sufficiently tightly constrained by the data. This is indeed the case, as we will see in Section~\ref{results}.

\subsection{Supernova lensing}
\label{lensing}

Lensing connects in a fruitful way the statistics of the matter distribution to the statistics of the luminosity distribution: by studying the latter one can gain information on the former and thus on the properties of matter clustering. In other words, one can use the scatter of the supernova magnitudes in the Hubble diagram to infer background and perturbation cosmological parameters.

In order to extract this lensing information, one must know how the lensing probability density function (PDF) depends on cosmology. To that end we employ the \tgl code, which is the numerical implementation of the (semi-analytical) stochastic gravitational lensing (sGL) method introduced in \citealt{Kainulainen:2009dw,Kainulainen:2010at,Kainulainen:2011zx}. In particular, the matter density contrast $\delta_{m}(r,t)$ is described as a random collection of halos, which we model according to the Navarro-Frenk-White (NFW) profile~\citep{Navarro:1995iw}. This is actually a one-parameter family of profiles, the parameter being the concentration parameter. Linear correlations in the halo positions are neglected: this should be a good approximation as the contribution of the 2-halo term is negligible with respect to the contribution of the 1-halo term~\citep{Kainulainen:2011zx}. Therefore, our matter modeling relies on a collection of NFW halos, whose properties are determined once a halo mass function $f(M,z)$ and a concentration parameter function $c(M,z)$ are provided. The observational determination of the former is the subject of this work. The latter is instead assumed to take the functional form proposed in~\citet[][Table 1]{Duffy:2008pz}. This is justified as we are fixing the cosmological parameters (see previous discussion).

In order to compute the supernova lensing likelihood we adopt the ``Method of the Moments''~(MeMo), proposed in~\cite{Quartin:2013moa}. The MeMo approach works by parametrizing the lensing PDF by its first statistical moments, which can then be propagated into the moments of the final PDF and confronted with data. In more detail, a $\chi^2$ is built with the first moment $\mu_{1}'$ (which is independent of lensing due to photon number conservation) and the first three central moments $\{\mu_{2},\mu_{3},\mu_{4}\}$ (which we will collectively refer to simply as $\mu_{1-4}$) and compare them with the corresponding theoretical predictions. The theoretical predictions of the moments of the final PDF are obtained from the convolution of the lensing PDF with the intrinsic supernovae dispersion PDF. To wit:
\begin{align}
    \mu_{2} & \;\equiv\;\sigma_{{\rm tot}}^{2}
    \;=\;\sigma_{{\rm lens}}^{2}+\sigma_{\rm int}^{2}\,,\label{eq:mu2}\\
    \mu_{3} & \;=\;\mu_{3,{\rm lens}} + \threeint \,,\label{eq:mu3}\\
    \mu_{4} & \;=\;\mu_{4,{\rm lens}}+6\,\sigma_{{\rm lens}}^{2}\,\sigma_{\rm int}^{2} +  3 \, \sigma_{\rm int}^{4} + \fourint \,, \label{eq:mu4}
\end{align}
where $\mu_{1-4,\rm lens}$ are the lensing moments (obtained with \tgl -- see \citealt{Amendola:2013twa}) and $\{\sigma_{\rm int},\threeint,\fourint\}$ are the (\emph{a priori} unknown) intrinsic moments of the SN luminosity distribution, which we define as including also instrumental errors. The MeMo likelihood is then:
\begin{equation}\label{Lmom}
\begin{aligned}
    & L_{\rm{MeMo}}\,=\,\prod_{j}^{\rm bins}\frac{1}{(2\pi)^2 \sqrt{|\Sigma_{j}|}} \exp \left(-\frac{1}{2}\,\chi_{j}^{2}\right), \\
    & \chi_{j}^{2} \,=\, \big(\boldsymbol{\mu}-\boldsymbol{\mu}_{\rm{data}}\big)^t \,\Sigma_{j}^{-1}\, \big(\boldsymbol{\mu}-\boldsymbol{\mu}_{\rm{data}}\big), \\
    & \boldsymbol{\mu}\,\equiv\,\{ {\mu_{1}^{\prime}},\mu_2,\mu_3,\mu_4 \}\,,
\end{aligned}
\end{equation}
where the components of $\boldsymbol{\mu}_{\rm obs}(z_{j})$ are the moments inferred from the data, which -- for the forecasts -- we take to be $\boldsymbol{\mu}$ evaluated at the fiducial model and at redshift $z_{j}$. Although the covariance matrices $\,\Sigma_j\,$ depend in principle on cosmology, to a good approximation they can be built using the fiducial moments, as discussed in~\cite{Quartin:2013moa}. In this way, they no longer depend explicitly on cosmology; they depend only on $z$.

The MeMo has already been tested on real data in~\citet{Castro:2014oja,Castro:2015rrx,Macaulay:2016uwy}. It yielded cosmological constraints consistent with other probes,  although the precision is still small. It was shown that one can set $\fourint=0$ and assume that the supernova intrinsic distribution does not depend strongly on redshift, and so we make these assumptions also here.\footnote{Also, in~\citet{Amendola:2014yca} it was shown that marginalizing over $\fourint$ provided basically the same results.} Summarizing, $L_{\rm MeMo}$ depends on the 4 parameters~$\{a, p,\sigma_{\rm int}, \threeint \}$. The fiducial value we adopt for $\threeint$ is zero. The fiducial value for $\sigma_{\rm int}$ is discussed in the next Section.
The presence of a possible nonzero $\threeint$ means that we are not assuming the intrinsic supernova distribution to be Gaussian.

\begin{figure*}
\begin{centering}
    \includegraphics[width=.88\textwidth ]{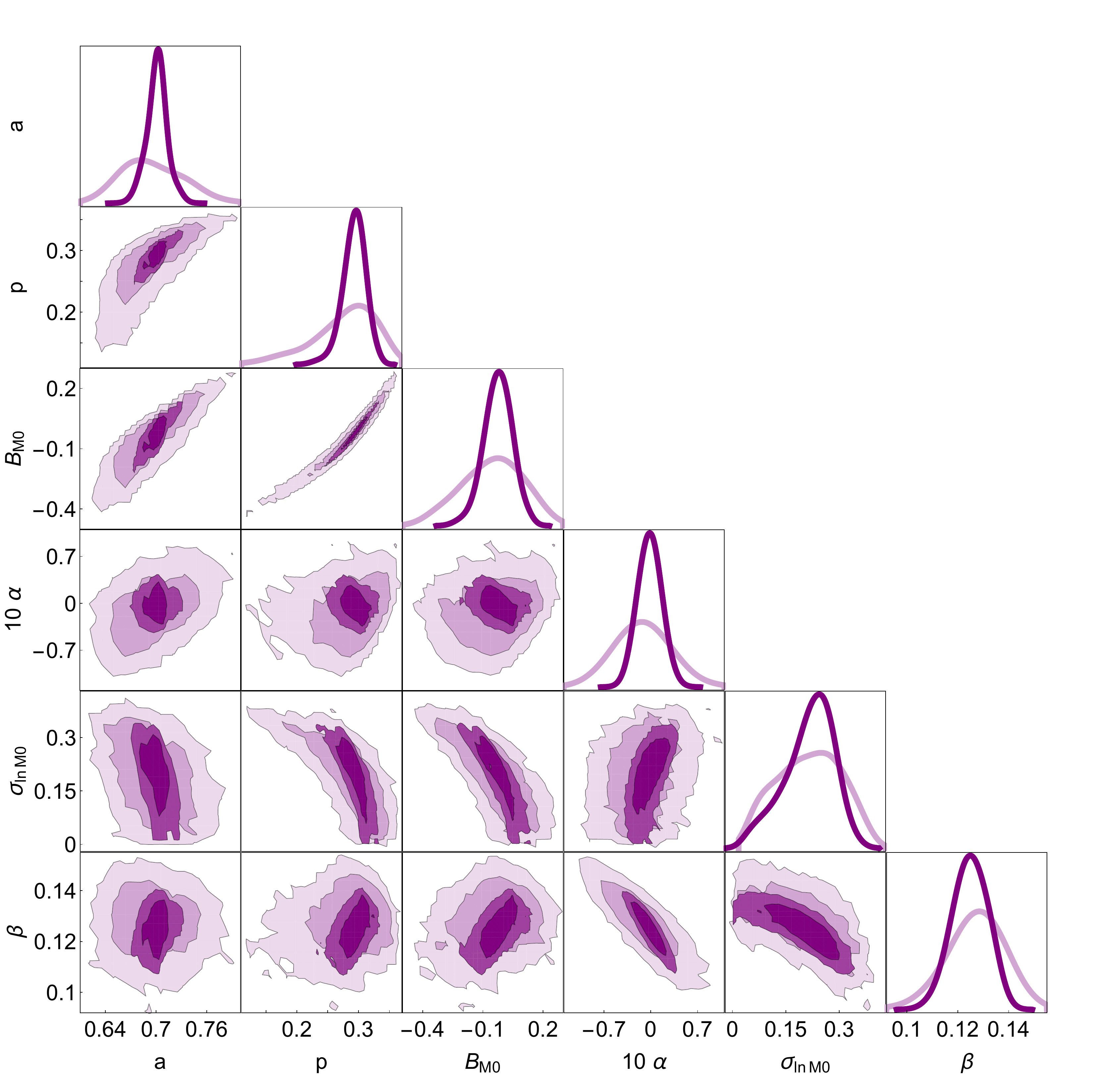}
    \caption{\label{fig:cctri}
    1- and 2$\sigma$ marginalized constraints on the relevant parameters of the cluster counts likelihood $L_{\rm cc}$ of equation~\eqref{Lcc} combined with the cluster power spectrum likelihood $L_{\rm cps}$ of equation~\eqref{likepk} for the forecasted Euclid clusters~\citep{Sartoris:2015aga}. The darker (lighter) contours represent the case of a S/N detection threshold of 3 (5) (see Figure~\ref{mthr}). See also Table~\ref{tab:ap} and Figure~\ref{figure:hmf}.}
\end{centering}
\end{figure*}

\subsubsection{Supernova data}
\label{cats}

The Dark Energy Survey is expected to observe over 3000 SNe during its observational cycle. We here assume a catalog built using the ``hybrid-5'' strategy discussed in~\citet{Bernstein:2011zf}. We also assume a total of 3000 observed SNe, normalizing the histogram plotted in Figure 10 of the latter paper. 
\citet{Bernstein:2011zf} also estimated in their Table 14 the final expected scatter in the Hubble diagram. The scatter varies non-linearly with $z$, with a minimum of 0.14 and a maximum of 0.25 mag. Although these numbers are related to the slightly different ``hybrid-10'' survey strategy, we assume here that they remain unchanged in ``hybrid-5''. Finally, we note that~\citet{Scovacricchi:2015ely} showed that an additional systematic error of 0.10 mag seems to have been implicitly assumed by~\citet{Bernstein:2011zf} in order to produce  their forecasts, but we neglect this small correction here.

The Large Synoptic Survey Telescope (LSST, see \citealt{Abell:2009aa}) is a wide-field photometric survey, currently under construction, that is expected to begin full operations in 2022. By the end of its ten-year mission the number of observed supernovae will be a few millions. This number includes all the expected observed supernovae but here we adopt the distribution based on the selection cut of signal-to-noise ratio higher than fifteen in at least two filters. The total number of supernovae decreases then to half a million in five years (we include SNe from both its ``main'' and ``deep'' surveys).
The dispersion in the Hubble diagram of the LSST SN catalog is not yet well understood. Rough estimations in the LSST white paper make it seem that a dispersion of 0.15 mag constant in redshift may be a reasonable hypothesis. On the other hand, as discussed above, DES, also a photometric survey, will apparently have a reasonably larger scatter. Therefore, we will consider two cases for LSST: $\sigma_{\rm int} = 0.15$ mag and $\sigma_{\rm int} = 0.20$ mag. Note that since we define $\sigma_{\rm int}$ to include noise, it corresponds to the total final Hubble diagram dispersion, which accounts for photometric redshift and other instrumental errors.

\section{Results}\label{results}

We will now present the forecasted constraints on the mass function parameters from Euclid, \jpas and DES. The cluster count and cluster power spectrum likelihoods depend on the six parameters~$\{a, \, p,\,B_{M0}, \,\alpha,\,\sigma_{\text{ln}M0}, \,\beta  \}$, while the SN lensing likelihood depends on the four parameters~$\{a, p,\sigma_{\rm int}, \threeint \}$.
While we will always consider flat priors on the intrinsic moments of the SN luminosity distribution as they cannot be independently measured, the parameters that enter the mass-observable relation could be calibrated with external X-ray data or weak lensing data and/or through cosmological simulations. Therefore, we start by adopting in the following Sections \ref{Euclid}-\ref{DES} very conservative flat priors on the mass-observable parameters, but in Section~\ref{perfect} we consider the other extreme case, to wit the one in which we have perfect knowledge of the mass-observable parameters.

\subsection{Euclid and LSST constraints}
\label{Euclid}

\begin{figure*}
\begin{centering}
\includegraphics[width=.42\textwidth]{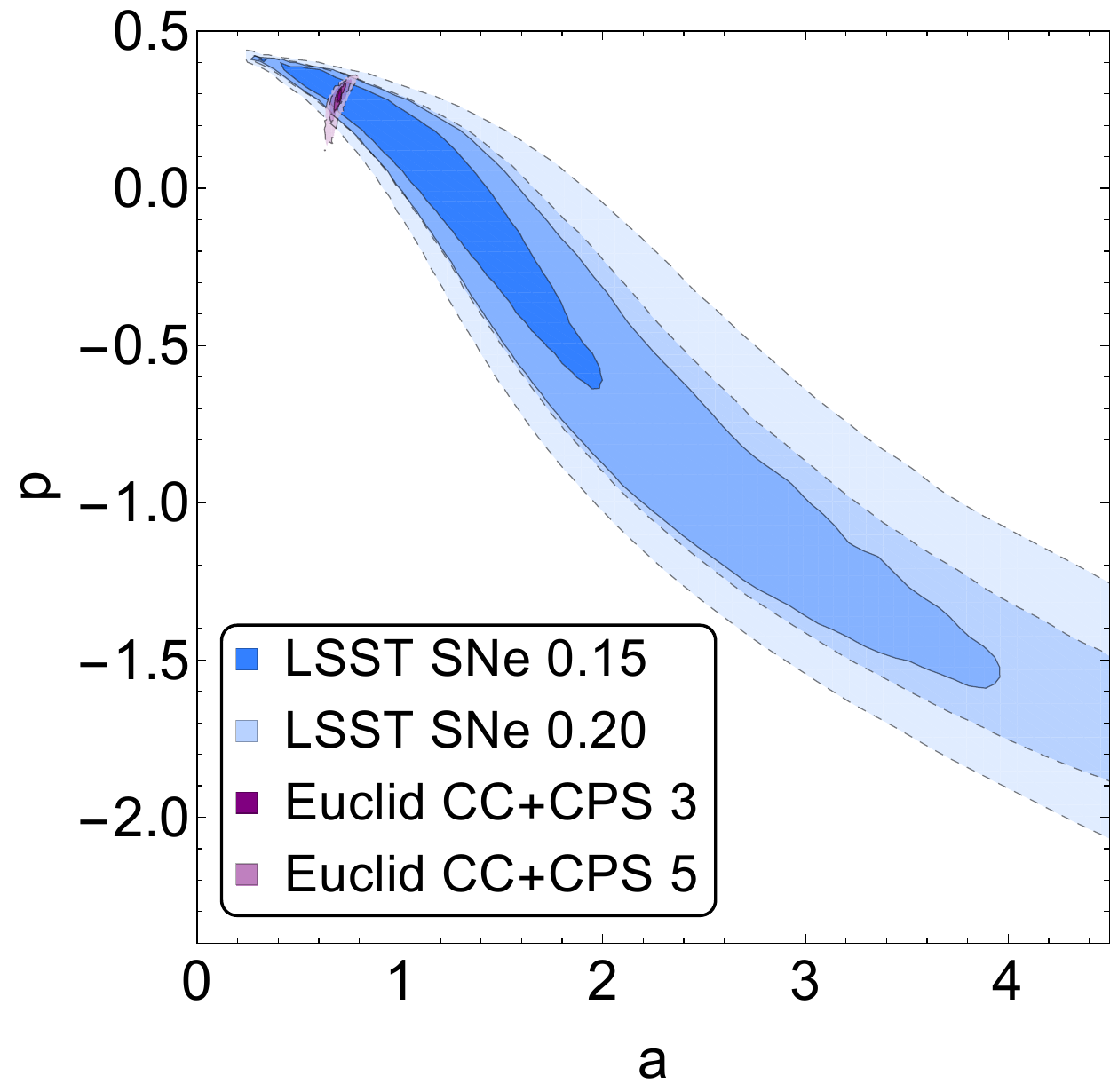}\qquad\quad\qquad
\includegraphics[width=.42\textwidth]{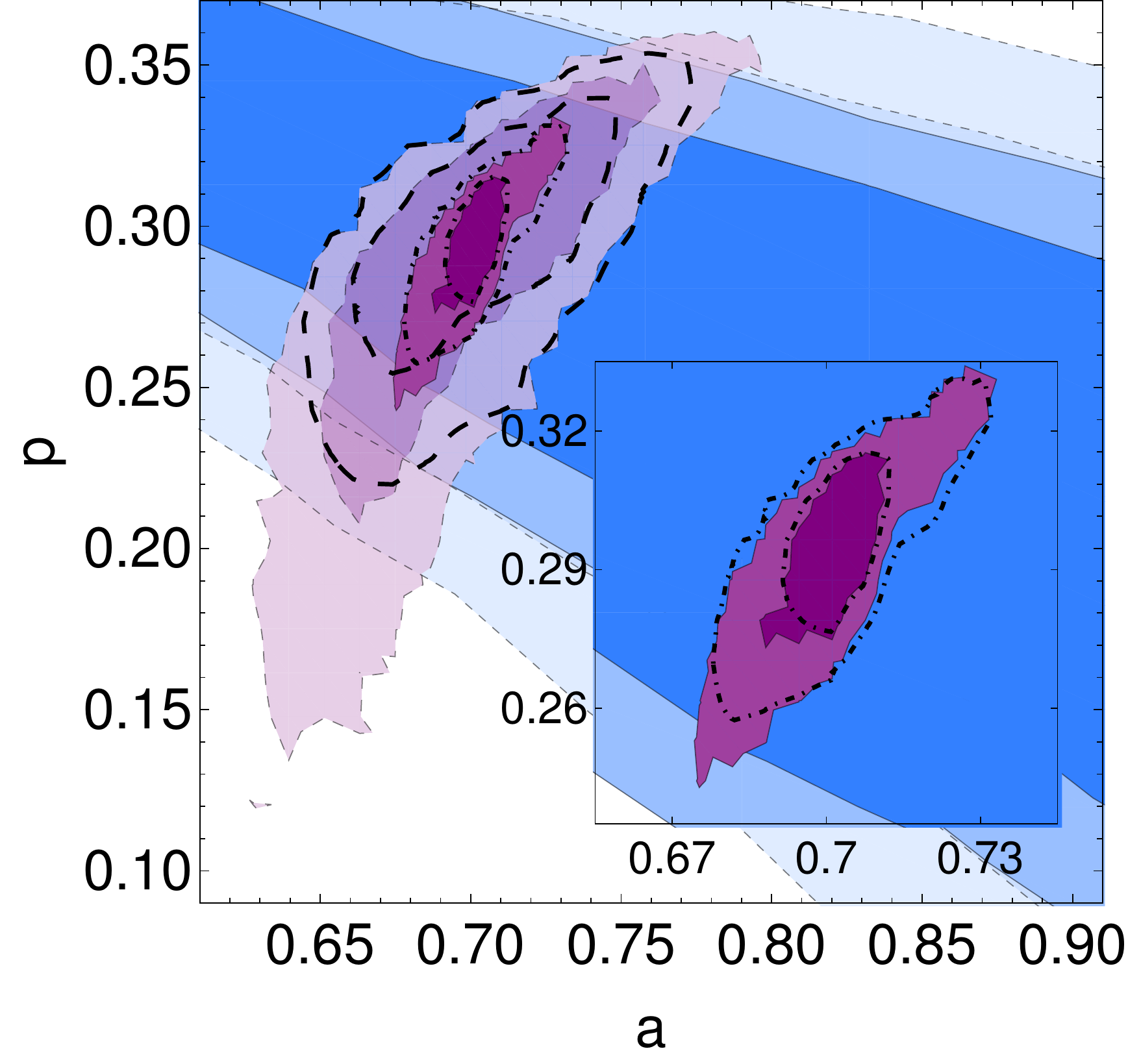}
\caption{1- and 2$\sigma$ marginalized constraints on the parameters $a$ and $p$ of the Sheth-Tormen mass function of eq.~\eqref{st}--the right panel is a zoom of the left panel. Darker (lighter) purple contours are relative to the cluster counts likelihood $L_{\rm cc}$ of eq.~\eqref{Lcc} combined with the cluster power spectrum likelihood $L_{\rm cps}$ of eq.~\eqref{likepk} for the Euclid clusters with a S/N detection threshold of 3 (5). See also Figure~\ref{fig:cctri}. Darker (lighter) blue contours are for the LSST supernova lensing likelihood $L_{\rm MeMo}$ of eq.~\eqref{Lmom} supposing  an intrinsic scatter of 0.15 (0.20) mag.
Dot-dashed (dashed) contours show the combined constraints of the analysis Euclid CC+CPS 3 + LSST SNe 0.15 (Euclid CC+CPS 5 + LSST SNe 0.20).
See also Table~\ref{tab:ap}.
\label{figure:hmf}}
\end{centering}
\end{figure*}

Figure~\ref{fig:cctri} shows marginalized 1- and $2\sigma$ constraints and correlations on the parameters $a$ and $p$ of the Sheth-Tormen mass function of equation \eqref{st} and on the four nuisance parameters of equations \eqref{Mbias}--\eqref{sigln} for the forecasted Euclid clusters~\citep{Sartoris:2015aga} using the  cluster count likelihood $L_{\rm cc}$ of equation~\eqref{Lcc} combined with the cluster power spectrum likelihood $L_{\rm cps}$ of equation~\eqref{likepk}.
Darker (lighter) colors refer to a detection threshold of 3 (5), see Figure~\ref{mthr}.

As discussed earlier, cosmological parameters are kept fixed to the best-fit values from Planck~\citep[][table 4, last column]{Ade:2015xua} since $a$ and $p$ inherit the universality of the mass function. The strong degeneracies shown in Figure~\ref{fig:cctri} among $\,a,\,p\,$ and the nuisance parameters emphasize the importance of properly modeling systematic uncertainties. From the marginalized 1$\sigma$ constraints on $a$ and $p$, shown in Table~\ref{tab:ap}, one concludes that Euclid will be able to constrain the halo mass function, although its constraining power is diminished by the degeneracy between $a$ and $p$.

In Figure~\ref{figure:hmf} the blue 1- and $2\sigma$ contours depict the marginalized constraints on the parameters $a$ and $p$ from the supernova lensing likelihood of equation~\eqref{Lmom}. Darker contours refer to the more aggressive assumption of $\sint=0.15$ mag, while brighter ones to the more conservative assumption $\sint=0.20$ mag. The constraints from the combination of cluster counts and cluster power spectrum likelihoods of Figure~\ref{fig:cctri} are shown in purple in Figure~\ref{figure:hmf}. Darker (brighter) purple contours refer to the aggressive (conservative) S/N threshold of 3 (5), as discussed in Section~\ref{clusters2}.

In both cases, the overall constraints from SN lensing are very weak. However, the degeneracy between $a$ and $p$ is approximately orthogonal with respect to the one featured by the cluster observables. Therefore, supernova lensing is able to improve somewhat Euclid's constraints on the mass-function parameter $p$ and -- to less extent -- on the parameter $a$ (see Table~\ref{tab:ap} for the numerical values). Once again \citep[see][]{Amendola:2014yca,Castro:2015rrx}, supernova lensing constraints have proven to suffer from different parameter degeneracies as compared to other more standard probes, thus efficiently complementing them. The same holds true as far as the robustness with respect to systematic uncertainties is concerned.

\subsection{\jpas cluster constraints}
\label{J-PAS}

\begin{figure}
\begin{centering}
\includegraphics[width=.85\columnwidth]{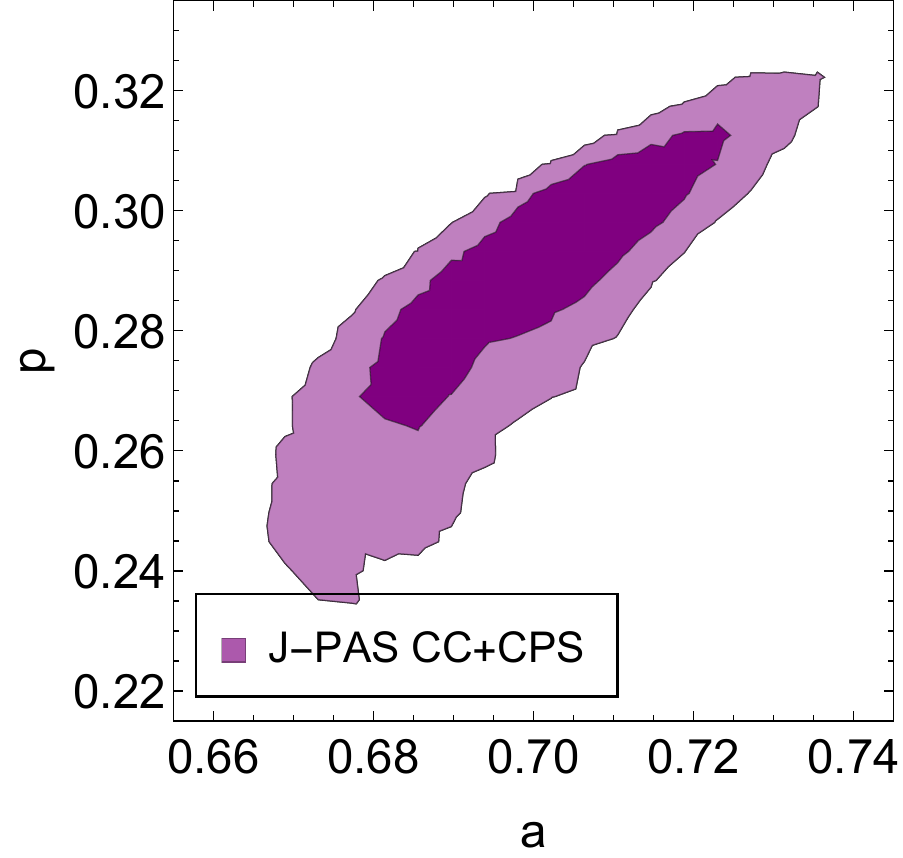}
\caption{
Same as Figure~\ref{figure:hmf} but for the forecasted \jpas clusters. See also Table~\ref{tab:ap}.
\label{figure:jpas}}
\end{centering}
\end{figure}

In Figure~\ref{figure:jpas} we repeat the analysis using the forecasted \jpas cluster catalog~\citep{Ascaso:2016ddl}.
The expected completeness in this catalog~\citep{Ascaso:2016ddl} is similar to the one in Euclid with the aggressive S/N threshold of 3. Interestingly, because of its lower mass-threshold \jpas should be able to perform almost as well as Euclid with the more aggressive threshold. That is, the presence of less-massive halos in \jpas for $z<0.75$ compensates for the smaller survey area and shallower depth.
See Table~\ref{tab:ap} for the marginalized constraints on the mass function parameters.

\subsection{DES cluster and supernova constraints}
\label{DES}

\begin{figure}
\begin{centering}
\includegraphics[width=.85\columnwidth]{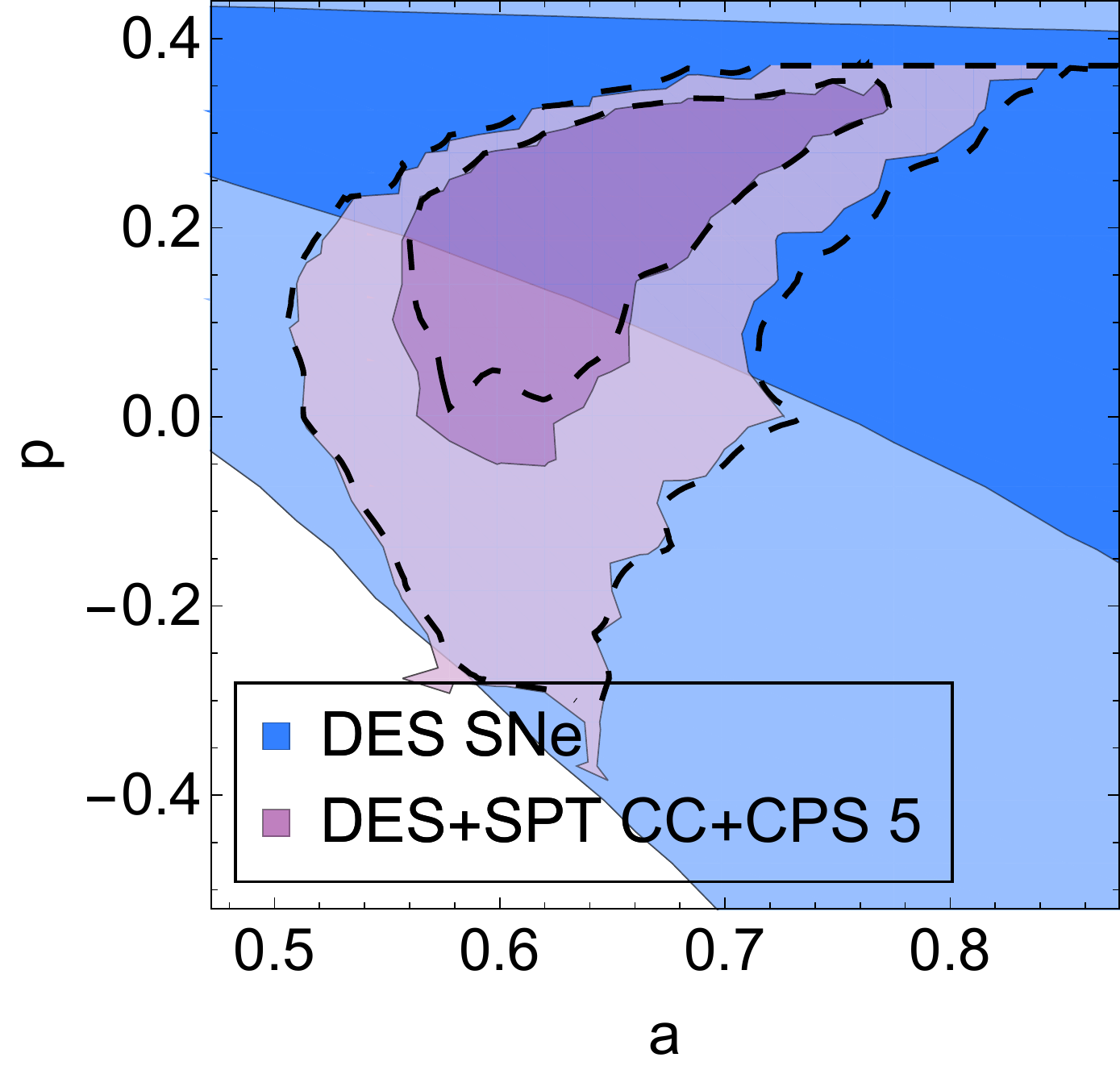}
    \caption{
    Same as Figure~\ref{figure:hmf} but for the forecasted DES clusters and supernova lensing. 
    See also Table~\ref{tab:ap}.
    \label{figure:hmf2} }
\end{centering}
\end{figure}

In Figure~\ref{figure:hmf2} we repeat the analysis using the forecasted supernova (blue contours) and cluster (purple contours) catalogs of DES+SPT with S/N detection threshold of 5. This combined survey is forecasted to observe about 9 times less clusters than Euclid with the same detection threshold of 5. The naive expectation of $\,\sim$3 times worse constraints is roughly confirmed by Figure~\ref{figure:hmf2}. In this case the SN lensing improvements are smaller and its main quality is to serve as a cross-check of systematics. See Table~\ref{tab:ap} for the marginalized constraints on the mass function parameters.

\subsection{Constraints with perfect knowledge of mass-observable relation}
\label{perfect}

In the previous Sections we have taken the very conservative assumption that we have no prior information on the nuisance parameters of equations \eqref{Mbias}--\eqref{sigln}. However, this will likely not be the case. For example, external X-ray data could be used to calibrate the cluster mass, as well as Sunyaev-Zel'dovich data or velocity dispersion counts~\citep{Caldwell:2016owp}. Alternatively, the weak-lensing signal in shear maps could be used to determine the true mass in a much more robust way. Finally, simulations may give priors on the nuisance parameters which are less penalizing than the flat priors adopted in the previous analysis. Therefore, it is important to consider also the other extreme case, in which we assume a perfect knowledge of the scaling relation between the observed and true galaxy cluster mass. The two cases we consider -- zero and perfect knowledge -- should bracket the range of possible constraints.

\begin{figure}
\begin{centering}
\includegraphics[width=.85\columnwidth]{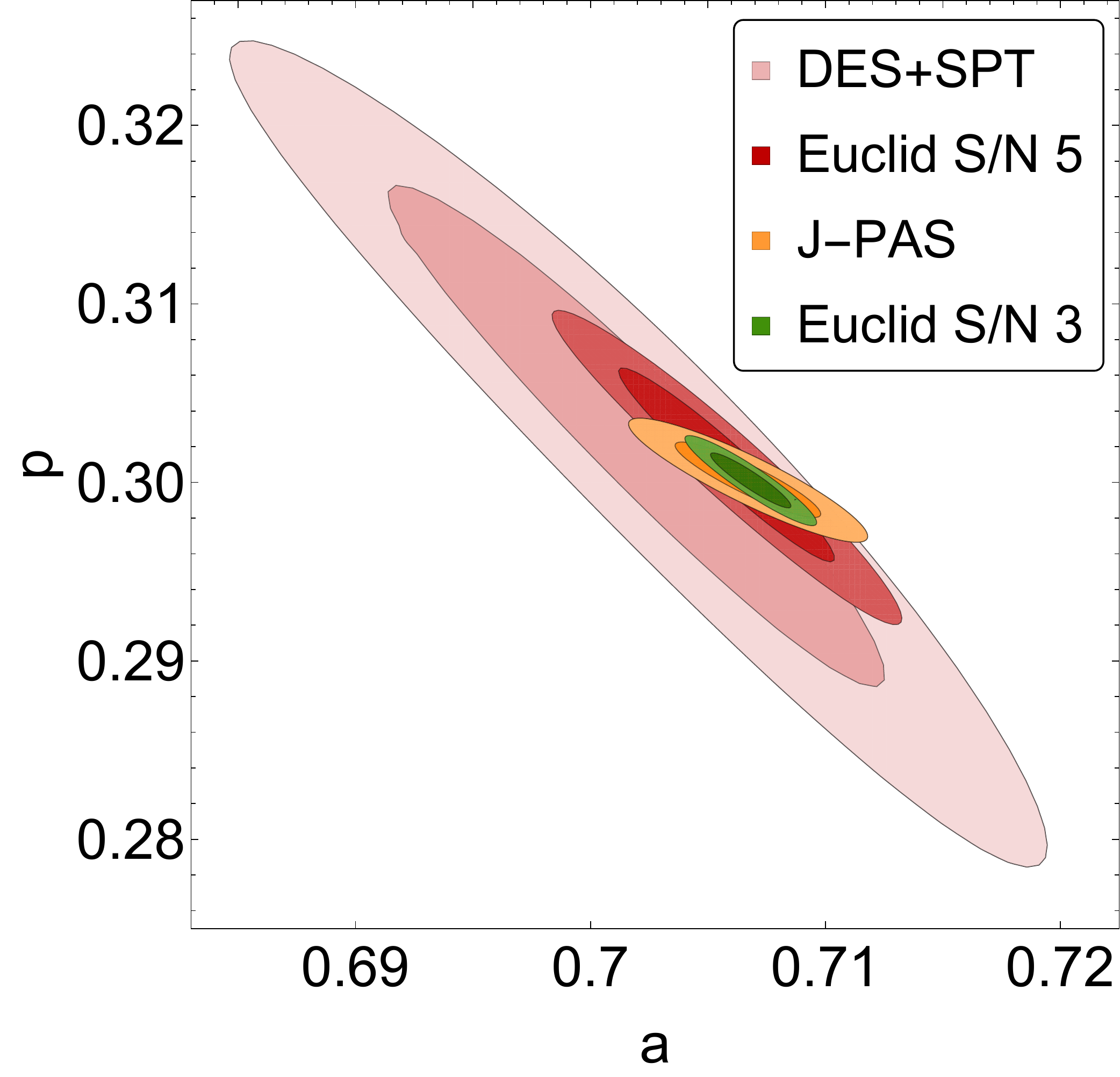}
\caption{
1- and 2$\sigma$ marginalized constraints on the parameters $a$ and $p$ of the Sheth-Tormen mass function of eq.~\eqref{st} using cluster counts and the cluster spectrum. Differently from Figures~\ref{figure:hmf}-\ref{figure:hmf2} here we assume a \emph{perfect knowledge of the mass-observable relation}.
See Section~\ref{perfect} for more details. See also Table~\ref{tab:ap}.
}
\label{fig:perfect}
\end{centering}
\end{figure}

In Figure~\ref{fig:perfect} we show the constraints on the mass function parameters for Euclid, \jpas and DES+SPT for the case in which the nuisance parameters are fixed to their fiducial value. Also, we restrict ourselves here to the CC+CPS data as SN data do not add any further information. As can be seen from Table~\ref{tab:ap}, complete knowledge of the scaling relations allows one to improve the constrains on \emph{each} parameter by a factor of 10. For comparison,~\citet{Sartoris:2015aga} found that similar improvements on cosmological constraints are smaller, roughly a factor of 2 or 3. This makes it clear that using external data to calibrate the mass-observable relations is crucial if one aims to improve the precision on measurements of the halo mass function. Note also that the main degeneracy direction between $a$ and $p$ rotates substantially with knowledge of the mass-observable relation. This further confirms the degeneracies shown in Figure~\ref{fig:cctri} (specially with the mass bias $B_{M0}$), and explains the large improvement in the constraints.

\begin{table}
\begin{center}
    \begin{tabular}{lll}
    \hline
    \hline
    \multicolumn{3}{|l|}{FLAT PRIOR ON MASS-OBSERVABLE RELATION}\\
    \hline
    data \hspace{4.4cm} $\,$& $\left<\sigma_a\right>\;\;$ & $\left<\sigma_p\right>$ \\
    \hline
    DES+SPT CC+CPS 5 &$0.059$  &$0.15 $ \\[5 pt]
    DES+SPT CC+CPS 5 + DES SNe  &$0.059$  &$ 0.13 $ \\[5 pt]
    \jpas CC+CPS & $0.017$  &$ 0.019 $ \\[5 pt]
    Euclid CC+CPS 5 & $ 0.036$ &$ 0.047$ \\[5 pt]
    Euclid CC+CPS 5 + LSST SNe 0.20 &$ 0.030$  &$ 0.028 $ \\[5 pt]
    Euclid CC+CPS 3 & $0.012$ &$ 0.017$ \\[5 pt]
    Euclid CC+CPS 3 + LSST SNe 0.15 &$ 0.0093$  &$ 0.012 $ \\
    \hline
    \hline
    \multicolumn{3}{|l|}{PERFECT KNOWLEDGE OF MASS-OBS.~RELATION}\\
    \hline
    data \hspace{4.4cm} $\,$& $\left<\sigma_a\right>\;\;\;$ & $\left<\sigma_p\right>$ \\
    \hline
    DES+SPT CC+CPS 5 &$ 0.0067 $  & $ 0.0092 $ \\[5 pt]
    \jpas CC+CPS & $ 0.0020 $  &$ 0.0014 $ \\[5 pt]
    Euclid CC+CPS 5 & $ 0.0030 $ &$ 0.0036 $ \\[5 pt]
    Euclid CC+CPS 3 & $0.0011$ &$ 0.0010 $ \\
    \hline
    \hline
    \end{tabular}
\end{center}
\vspace{-.1cm}
\caption{
Top: Constraints on the parameters $a$ and $p$ of the Sheth-Tormen HMF of eq.~\eqref{st} with a flat prior on the nuisance parameters of eqs.~\eqref{Mbias}--\eqref{sigln}.
Bottom: Same but with perfect knowledge of the nuisance parameters.
$\left<\sigma_X\right>$ stand for the average 1$\sigma$ uncertainty in $X$, \emph{i.e.}, $(\sigma^+_X + \sigma^-_X)/2$; ``CC+CPS $X$''  for the combined analysis of cluster counts and  power spectrum with S/N threshold of $X$; ``SN $Y$'' for an assumed final scatter in the SN Hubble diagram of $Y$ mag. See Figures~\ref{figure:hmf}-\ref{fig:perfect} for the corresponding 2D posteriors. The fiducial values used were $a=0.707$, $p=0.3$.
\label{tab:ap}}
\end{table}%

\subsection{Root mean square residuals}
\label{rms}

Root mean square (rms) residuals are a useful indicator for deviations in the mass function.
We define the rms residuals according to:
\begin{equation} \label{eq:rms}
{\rm rms} = \sqrt{\langle (f_j/f_{j, {\rm fid}}-1)^2  \rangle} \,,
\end{equation}
where
\begin{equation}
f_j = \int_{\Delta M_j} \rd M \, n(M, z)
\end{equation}
is the binned halo count in $\Delta M_j=M_{j+1}-M_j$, and the fiducial values given in \eqref{stfid} have been used to calculate $f_{j, {\rm fid}}$.

\begin{figure}
\begin{centering}
\includegraphics[width=.85\columnwidth]{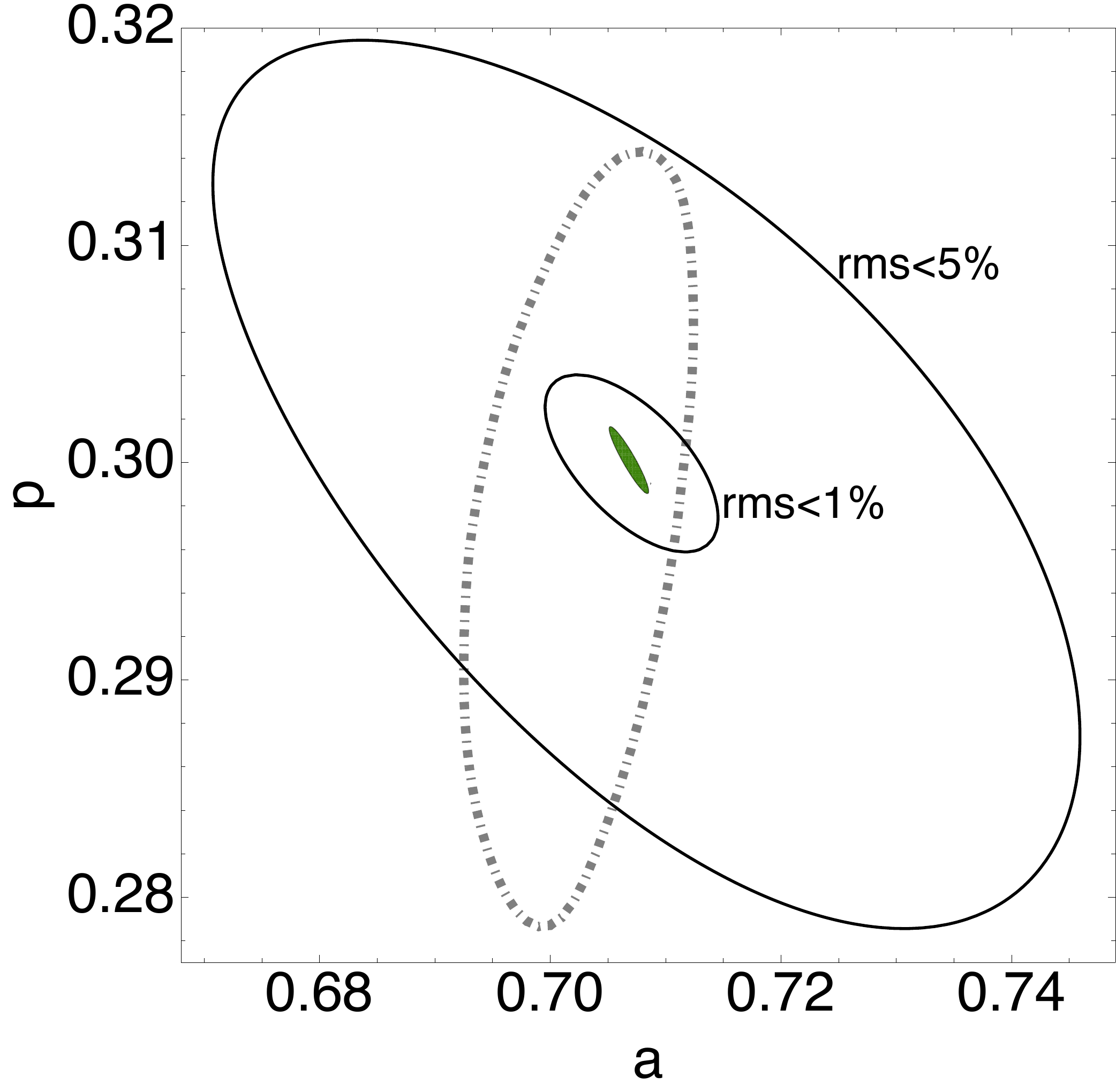}
\caption{
Regions bounding a given level (1\% and 5\%) of root mean square (rms) residuals in the mass function at $z=0$; see equation \eqref{eq:rms}.
The dot-dashed contour shows the 1-$\sigma$ constraints from Euclid CC+CPS 3 with a flat prior on the nuisance parameters; it is a smoothed version of the contour shown in Figure~\ref{figure:hmf}.
The small green contour shows the 1-$\sigma$ constraint from Euclid CC+CPS 3 in the case of a perfect knowledge of the mass observable relation (also shown in Figure~\ref{fig:perfect}).
See Section~\ref{rms} for more details.
}
\label{fig:rms}
\end{centering}
\end{figure}

In figure~\ref{fig:rms} we show the parameter space region where rms<1\% and rms<5\% at $z=0$. For the sake of comparison, we have overlapped the 1-$\sigma$ confidence regions relative to the most aggressive case of Euclid CC+CPS 3, for the two cases of a flat prior on the nuisance parameters and their perfect knowledge.
From this analysis we can conclude that  Euclid CC+CPS 3 with a flat prior on the nuisance parameters will be able to constrain the mass function to a 5\% level precision.
Instead, if we assume perfect knowledge of the nuisance parameters the constraints will be significantly better than 1\%.

\section{Conclusions}\label{conclusions}

In this work we discuss how observations can constrain the halo mass function. To our knowledge, this has not been studied before. As argued in the Introduction this is a sensible question to ask as halo mass functions from simulations may suffer from systematic effects, specially when baryonic (or any sub-grid) physics is included. There may even be physics beyond the standard model which has not yet been considered in simulations. In both cases, these effects may manifest themselves in the future as tensions between the observed and simulated halo mass functions.

Our results show that, in absence of external priors on the parameters of the mass-observable relation, future data from the Euclid Mission and \jpas can yield constraints on the mass function parameters which are comparable -- in the most favorable case -- to the ones from state-of-the-art simulations. Constraints from the Dark Energy Survey will be a factor of 3 worse than the above ones, providing therefore only weak constraints on the mass function parameters. By a rough estimation based on catalog sizes one concludes that present-day cluster data from Planck~\citep{Ade:2015fva} and from the South Pole Telescope~\citep{Bleem:2014iim} and supernova JLA data from~\citet{Betoule:2014frx} should perform according to a further factor of 3 worse as compared to DES. We therefore conclude that future datasets are needed in order to obtain valuable constraints.

The constraints above can potentially substantially improve with a better knowledge of the mass-observable relation. For example, external X-ray data and shear maps could be used to calibrate the cluster mass, and simulations may give priors on the nuisance parameters which are less penalizing than the flat (ignorance) priors used above. Therefore, it is worth considering the case of a perfect knowledge of the scaling relation between the observed and true galaxy cluster mass. In this case the constraints on each of the mass function parameters improve by a factor of roughly 10. Therefore, this preliminary analysis suggests that it is not only possible to test the physical mechanisms behind the halo mass function but also to achieve percent level accuracy. The two cases considered -- zero and perfect knowledge -- should bracket the range of possible constraints.

Future data could thus, at the same time, help calibrate large cosmological simulations which include hard-to-model baryonic physics and hint for physics beyond the standard model, if deviations are detected that cannot be accounted for. Indeed, from the analysis of Figure~\ref{stplot} we can conclude that future data are expected to have the precision to probe the systematic differences between pure dark-matter simulations and also between dark-matter and hydrodynamical simulations.

It should be stressed, however, that these conclusions depend, to some extent, on the chosen Sheth-Tormen mass function template which does not perform optimally for high halo masses. Ideally, one would want to constrain the mass function in a non-parametric way so as to make the analysis independent from modeling uncertainties. This will be subject of future research.

\section*{Acknowledgments}
It is a pleasure to thank Luca Amendola, Begoña Ascaso, Stefano Borgani, Luciano Casarini, Marcos Lima, and Barbara Sartoris for useful discussions. We also would like to warmly thank Alex Saro, Sebastian Bocquet and Klaus Dolag for sharing the halo catalogs of the Magneticum simulation, and Raul Angulo for making publicly available the catalogs of the Millennium-XXL simulation.
Part of the data postprocessing and storage has been done on the CINECA facility PICO.
TC is supported by the Brazilian research agency CAPES and
by a Science Without Borders fellowship from the Brazilian National Council
for Scientific and Technological Development (CNPq).
VM is supported by the Brazilian research agency CNPq.
MQ is supported by the Brazilian research agencies CNPq and FAPERJ.

\bibliography{growth-rate,halo,cosmo-lensing}

\appendix

\vspace{1cm}

\section{Fitting functions}\label{app:pfits}

Here we give analytical fitting functions for the second-to-fourth central lensing moments $\mu_{2-4, \text{lens}}$ as a function of redshift $z$ and the parameters $a$ and $p$ of the Sheth-Tormen mass function of equation \eqref{st}, which are valid within the domain:
\begin{align*}
    0 <  & \; z < 1.2 \,,   \\
    0.5 < & \; a  <   1.3  \,, \\
    0  < & \; p  <  0.4 \,.
\end{align*}
All the other cosmological parameters are fixed to the best-fit values from Planck~\citep[][table 4, last column]{Ade:2015xua}.
Using magnitudes, the fitting formulae are:
\begin{equation} \label{2fit}
\begin{aligned}
    \sigma_{\rm lens}(&z, a,p) \,=\,-\frac{0.003 z}{a^2}+0.0046 a p z+\frac{0.024 z}{a}+0.069 P z^3 \\
    & -0.076 P^2 z^2-0.18 P z^2-0.021 P^2 z-0.014 P z \, , \\
    \mu_{3,{\rm lens}}^{1/3}(&z, a,p) \,=\,-\frac{0.0082 z}{a^2}+0.013 a p z+\frac{0.046 z}{a}+0.081 P z^3\\
    &+0.042 P^2 z^2-0.094   P z^2-0.14 P^2 z-0.12 P z \,, \\
    \mu_{4,{\rm lens}}^{1/4}(&z, a,p) \,=\,-\frac{0.016 z}{a^2}+0.026 a p z+\frac{0.076 z}{a}+0.077 P z^3 \\
    &+0.23 P^2 z^2+0.053   P z^2-0.35 P^2 z-0.3 P z \,,
\end{aligned}
\end{equation}
where $P=p-1/2$. In the entire domain of validity, the average RMS error is 0.00083, 0.00096 and 0.0018 for $\mu_{2-4, \text{lens}}$, respectively, which is roughly 3\% for all three moments.

\label{lastpage}
\end{document}